\documentclass{article}
\usepackage{spconf,amsmath,graphicx,latexsym, amssymb, subcaption, }

\usepackage{enumitem}
\setlist{nosep, leftmargin=14pt}

\usepackage{mwe} % to get dummy images

\title{FocusNet++: Attentive Aggregated Transformations for Efficient and Accurate Medical Image Segmentation}
 \name{Chaitanya Kaul$^{\star}$ \thanks{Chaitanya Kaul and Roderick Murray-Smith acknowledge support from the iCAIRD project, funded by Innovate UK (project number 104690).} \quad Nick Pears$^{\dagger}$ \quad Hang Dai$^{\ddagger}$ \quad Roderick Murray-Smith$^{\star}$ \quad Suresh Manandhar$^{\dagger\dagger}$}

 \address{$^{\star}$ School of Computing Science, University of Glasgow, G12 8QQ, United Kingdom \\
     $^{\dagger}$ Department of Computer Science, University of York, YO10 5DD, United Kingdom \\
     $^{\ddagger}$ Mohamed bin Zayed University of Artificial Intelligence, Abu Dhabi, United Arab Emirates \\
     $^{\dagger\dagger}$ NAAMII, Katunje, Bhaktapur, Kathmandu, Nepal}

\begin{document}
%\ninept
%
\maketitle
\begin{abstract}
We propose a new residual block for convolutional neural networks and demonstrate its state-of-the-art performance in medical image segmentation. We combine attention mechanisms with group convolutions to create our group attention mechanism, which forms the fundamental building block of our network, FocusNet++. We employ a hybrid loss based on balanced cross entropy, Tversky loss and the adaptive logarithmic loss to enhance the performance along with fast convergence. Our results show that FocusNet++ achieves state-of-the-art results across various benchmark metrics for the ISIC 2018 melanoma segmentation and the cell nuclei segmentation datasets with fewer parameters and FLOPs.
\end{abstract}
\begin{keywords}
Group Attention, Medical Image Segmentation, Residual Learning
\end{keywords}
\section{Introduction}
\label{sec:intro}

Recently, the use of attention mechanisms in deep learning has been shown to learn better features \cite{focusnet} \cite{attentionunet}. Learning better feature extractors is the most important task a network can do, especially for attention-based architectures, as the attention mechanisms are learnt over the extracted features. This has resulted in an general emphasis on optimizing convolutions. The simplest form of attention networks are the Spatial Transformer Networks \cite{stn} that learn the regions of interest from images with random clutter or noise. One of the first major visual attention methods was a two-level approach \cite{twolevelattention}, where the images were first passed through an RCNN and selective search algorithms to generate proposals, followed by a gating operation using softmax over the ImageNet classes to remove low probability proposals. The remaining patches were then passed through a SVM classifier. The approach worked well on a subset of the ImageNet dataset, but requires a large amount of computation as well as hyperparameter tuning. SE-Nets \cite{se} proposed to global average pool feature map (channel) information into a single vector creating a global representation. Using 'Squeeze-and-excitation', CNNs levaraged channel wise context to improve accuracy. One of the first works to explicitly show how filter groups leads to learn better representations is {\it Deep roots} \cite{deeproots}, where, a sparse connecting structure resembling a tree root reduces parameters without any significant effect on the network accuracy. The impact of group convolutions was made apparent with ResNeXt \cite{resnext} which performed impressively on the ILSVRC 2016 tasks. In this research, we extend group convolutions by incorporating attention mechanisms inside filter groups. \\
To this end, we propose FocusNet++, a deep learning architecture for medical image segmentation that harnesses the power of grouped convolutions, and combines them with a FocusNet-style attention mechanism \cite{focusnet} to get an improved performance compared to FocusNet, with fewer than half the parameters than it's successor. We enhance the decoding using fine-grained information from each decoder scale, which helps improve the network's segmentation ability. We compare with state of the art architectures, namely, Wide UNet and UNet++ \cite{unet++}, R2U-Net \cite{recunet}, Attention U-Net \cite{attentionunet}, BCDU-Net \cite{bcdunet} and FocusNet \cite{focusnet} to show the superiority of our method. \\
The rest of this paper is organized as follows. We introduce FocusNet++ in Section \ref{focusnet++sec} where we describe our novel group attention block, as well as the loss function used for our experiments. Section \ref{experiments} summarizes our experiments on the skin cancer and cell nuclei segmentation datasets highlighting our model's state-of-the-art performance with reduced parameters and FLOPs compared to benchmark state-of-the-art medical image segmentation architectures. Our conclusions are provided in Section \ref{conclusion}.

\section{FocusNet++}
\label{focusnet++sec}

\begin{figure*}[htb]
\begin{center}
\includegraphics[scale=0.65]{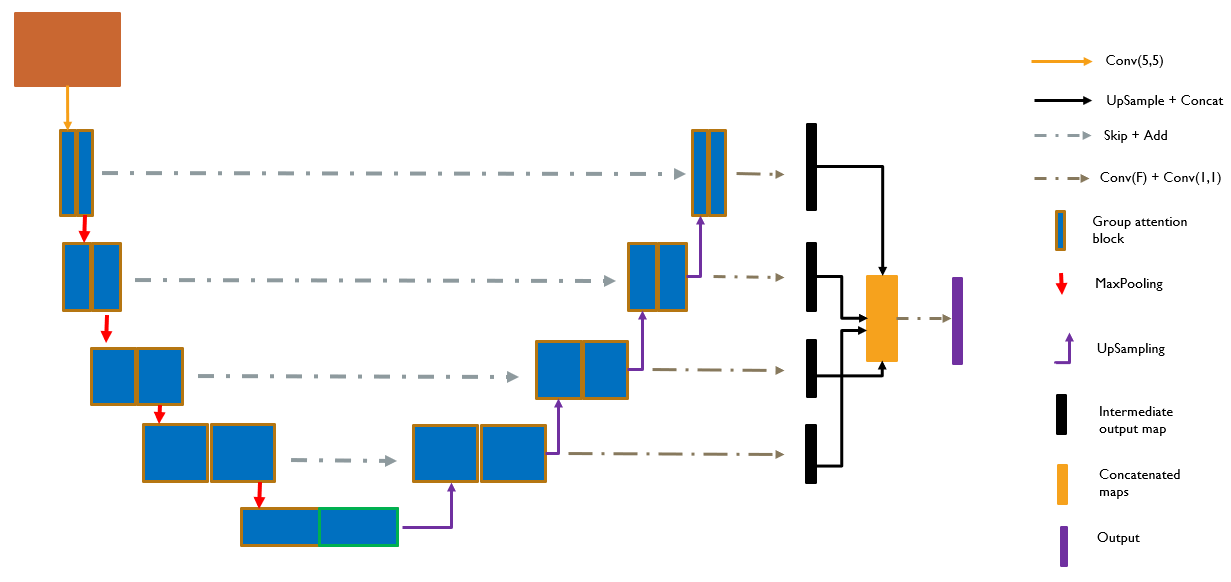}
\end{center}
\caption{The figure shows the architecture diagram for FocusNet++. The input image is processed by a series of residual group attention-max pooling blocks into a bottleneck and then decoded into a segmentation mask.}
\label{focusnet++}
\end{figure*}

Figure \ref{focusnet++} shows the FocusNet++ architecture that adopts an encoder-decoder structure to learn multi-scale features for medical image segmentation. We carefully designed a group attention block, called the Residual Group Attention Network, that employs our novel attention methodology inside group convolutions for effective feature learning. Our network aims to address the problem of the relatively inferior decoding ability of existing segmentation architectures, combined with better feature extraction via our group attention block. We do this by creating a scheme that combines the output from each decoder scale to the final output, which leads to superior performance. The output from each scale passes through a {\it Conv-BN-LeakyReLU-Conv-Sigmoid} block to give intermediate outputs that are up-sampled, if required, to the output size, and then concatenated. \\
Finally, the concatenated volume is passed through a {\it Conv-BN-LeakyReLU-Conv-sigmoid} block to get the output segmentation map. In the encoder, the feature is down-sampled by a max-pooling operation. We add skip connections from the encoder to the decoder, rather than concatenating them. To up-sample the feature in decoder, we repeat the values in a kernel from a lower scale into an up-sampling scale and letting convolutions learn their values. We use dropout operation in the bottleneck layer to avoid over-fitting. \\
The receptive field of the first convolution kernel is $5 \times 5$. Following that, all convolutions kernels have a receptive field of $3 \times 3$ when used for feature extraction, and $1 \times 1$ when used to learn attention weights (i.e. preceding the sigmoid gating). The number of filters in each layer are $32 \rightarrow 64 \rightarrow 128 \rightarrow 192 \rightarrow 256 \rightarrow 192 \rightarrow 128 \rightarrow 64 \rightarrow 32$, divided into 4 filter groups in each layer. The general structure of FocusNet++ is similar to the U-Net architecture, apart from the mentioned changes.

\subsection{Residual Group Attention Network (ResGANet)}
\label{attentionblock}

\begin{figure}[htb]
\begin{center}
\includegraphics[scale=0.5]{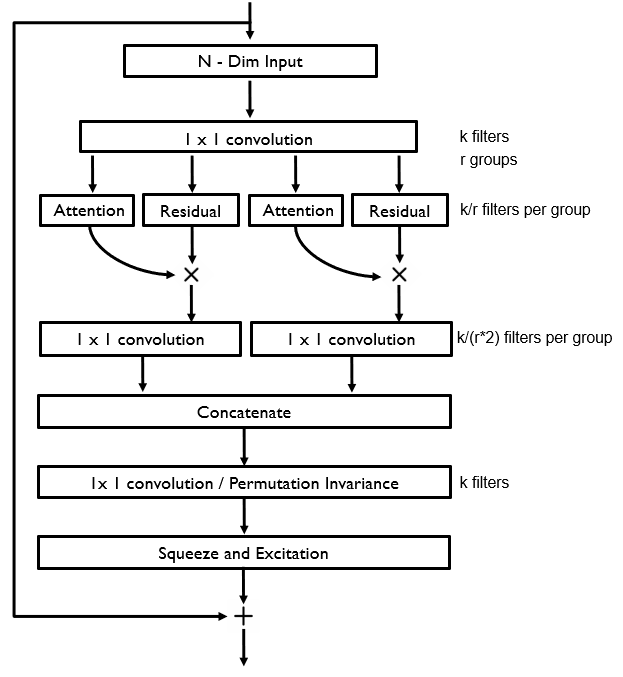}
\end{center}
\caption{Our novel residual block that first employs pixel-wise attention inside filter groups, followed by combining the groups via a permutation invariant embedding. The squeeze-and-excitation block then re-calibrates the feature maps, which is followed by the residual mapping.}
\label{groupattentionmodule}
\end{figure}

As shown in Figure \ref{groupattentionmodule}, the ResGANet employs pixel-wise attention inside filter groups, followed by combining the groups via a permutation invariant 1D convolution embedding. The squeeze and excitation block then re-calibrates the feature maps, which is followed by the residual mapping. The input to ResGANet is a feature volume that is processed by a $1 \times 1$ convolution operation. The general form of the aggregated transformation mapping is, $$ M = \complement_{i=1}^{r}P_i(x) $$ where $M$ is the output of the residual block, $\complement$ denotes concatenation, and $P_i(x)$ is some transformation learnt by $r$ separate stackings of trainable neurons transforming some input $x$. Here, $r=4$, as we divide this input features into groups of four, to be processed by four separate convolution groups. Each group, alternatively, is responsible for learning the attention weights for the group to it's right, and the next group learns the features that need to be extracted. The attention weights for each attention group are obtained via two {\it BN-LeakyReLU-Conv} operations followed by a {\it Conv-Sigmoid} operation to get the per-pixel weights. Each attention group transforms its input in the following way, $$ A_r = \sigma(\mathbf{W_a}, \delta(x_r, \mathbf{W_k)}) $$ where $\mathbf{W_k}$  and  $\mathbf{W_a}$ are the convolution weights and the attention weights respectively, $x_r$ is the $r^{th}$ group that is input into this block, and $\delta$ denotes the {\it LeakyReLU} activation. The residual block contains two {\it BN-LeakyReLU-Conv} operations followed by a skip connection that adds the features from the previous step to the residual block features. If the residual mapping is given by $O_r = x_r + F(x_r)$ then the network learns this $F(x_r)$ using some weights $\mathbf{W_k}$ as $F(x_r) = \delta(x_r, \mathbf{W_k})$. The output from the residual block is multiplied point-wise with the output from the attention block as $A = A_r \bigodot O_r $, weighting the pixels with a higher importance more prominently. Here, $\bigodot$ denotes the Hadamard product. The attention-infused output for each group propagates further in the block and is processed with a convolution block with a $1 \times 1$ receptive field and twice the number of filters. These intermediate filter maps are then concatenated and passed through a final $1 \times 1$ convolution. The feature maps are then re-calibrated using a squeeze-and-excitation operation that, finally, is followed by a residual connection.

\subsection{Hybrid adaptive logarithmic loss}
\label{loss}

In order to have better recall, we adapt the balanced cross entropy loss with the Tversky loss in a novel way to create our hybrid loss function. The loss is defined as, 
\begin{equation}
HL = (k)C_b + (1-k)TL
\end{equation}
where $C_b = \Omega p \log(\hat{p}) + (1 - \Omega)(1 - p) \log(1 - \hat{p}),$ $TL = \sum_{c} (1 - TI),$ the subscript indicates a summation over the number of classes $c$, and $TI = \frac{|G \cap P|}{|G \cap P| + \alpha|P \backslash G| + \beta|G \backslash P|}.$ 
To create a higher emphasis on the true positives, we select $\Omega = 0.7$. Generally, $\alpha = 0.3, \ \beta = 0.7$ proves to be the optimal setting in TL, adding higher weights to optimize over false positives and false negatives, so we retain those hyperparameter values. We weight the influence of both losses equally by setting $k=0.5$. In order to optimize the loss further, we use a function whose derivative gives a non-linear response closer to the global minimum leading to a heavy penalty for mis-classification. Hence, to mitigate the problem of pixel-class imbalance and poor convergence close to the minimum, we use the adaptive logarithmic loss \cite{all} for our problem. The loss is defined as,
\begin{equation}
	ALL-HL(x)=
	\left\{ \begin{array}{ll}
	 	\omega \ln\left(1 + \frac{|HL|}{\epsilon}\right) &|HL| < \gamma \\
	 	|HL|-C &\texttt{otherwise}
	 \end{array} \right.
\end{equation}
where $C = \gamma - \omega \ln\left(1+ \left(\frac{\gamma}{\epsilon}\right)\right)$. We observe that the default hyperparameters of this loss are optimal for our experiments. Hence, we set $\gamma = 0.1, \ \omega = 10.0$ and $ \epsilon = 0.5$.

\begin{figure*}[htb]
\begin{center}
\includegraphics[width=0.85\linewidth]{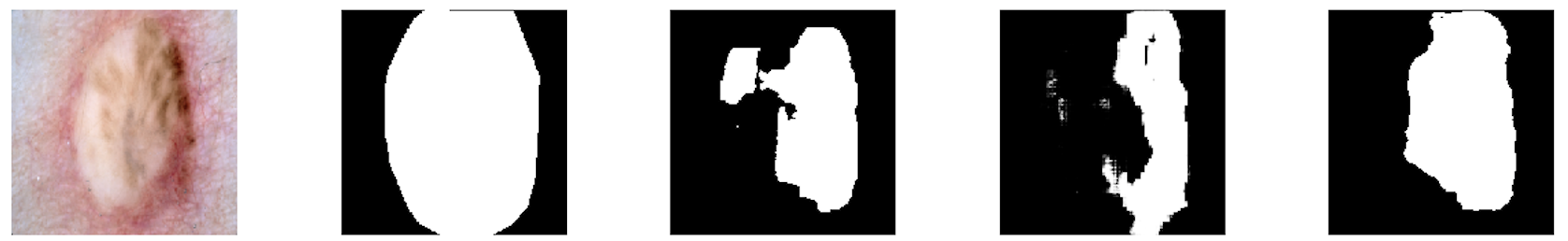}
\end{center}
\caption{Results for skin cancer segmentation. From left, original image, ground truth, segmentation results from Attention U-Net \cite{attentionunet}, FocusNet \cite{focusnet} and FocusNet++.}
\label{isicseg}
\end{figure*}

\begin{figure*}[htb]
\begin{center}
\includegraphics[width=0.85\linewidth]{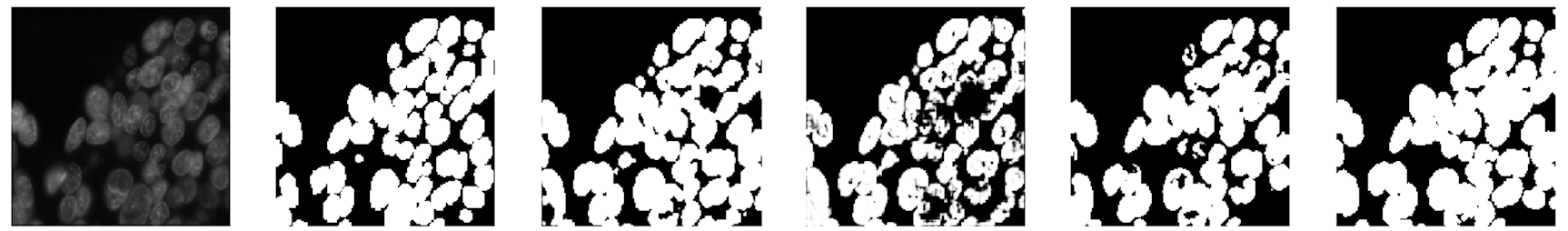}
\end{center}
\caption{Results for cell nuclei segmentation. From left, original image, ground truth, segmentation results from BCDU-Net \cite{bcdunet}, Attention U-Net \cite{attentionunet}, FocusNet \cite{focusnet} and FocusNet++.}
\label{nucleiseg}
\end{figure*}

\section{Experiments}
\label{experiments}

For all experiments, the train-validate-test data split is fixed and no data augmentation is used. As a pre-processing step, we scale all pixel values to the range [0,1]. We convert the segmentation mask to binary by setting every pixel above the threshold of 0.5 to 1. All our experiments are trained with the hybrid loss (ALL-HL) strategy. The experiments are conducted using Keras \cite{keras} using a TensorFlow backend. The batch size for all experiments is kept constant at eight. The networks are trained on Nvidia GTX 1080Ti GPUs using a carefully constructed learning rate schedule, optimized for every architecture. All architectures were trained for a maximum of 50 epochs and the best model weights were saved by monitoring the validation loss. No early stopping was used.

\subsection{Skin Cancer Segmentation}
\label{skincancer}

\begin{table}[htb]
\begin{center}
\begin{tabular}{|c||c|c|c|c|c|c|}
\hline
Method & Precision & Recall & DI & JI\\
\hline
\hline
FCN \cite{fcn} & 0.7176 & 0.8966 & 0.7861 & 0.7013\\
\hline
U-Net \cite{unet} & 0.7398 & 0.9043 & 0.8167 & 0.7268\\
\hline
Wide UNet \cite{unet++} & 0.7439 & 0.9167 & 0.8224 & 0.7334\\
\hline
R2U-Net \cite{recunet} & 0.7381 & 0.9122 & 0.8271 & 0.7511\\
\hline
BCU-Net \cite{bcdunet} & 0.7576 & 0.9272 & 0.8637 & 0.7665\\
\hline
UNet++ \cite{unet++} & 0.7516 & 0.8889 & 0.8437 & 0.7435\\
\hline
Attn U-Net \cite{attentionunet} & 0.7526 & 0.9286 & 0.8741 & 0.7813\\
\hline
FocusNet \cite{focusnet} & 0.7805 & 0.9328 & 0.8676 & 0.7751\\
\hline
FocusNet++ & \textbf{0.8322} & \textbf{0.9471} & \textbf{0.9014} & \textbf{0.8271}\\
\hline
\end{tabular}
\end{center}
\caption{Segmentation results on ISIC 2018 dataset.}
\label{isic2018falpha}
\end{table}

\begin{table}[htb]
\begin{center}
\begin{tabular}{|c||c|c|c|}
\hline
Architecture & Params & FLOPs \\
\hline
\hline
UNet \cite{unet} & 7.94$\times 10^8$ & 16.12$\times 10^8$\\
\hline
UNet++ \cite{unet++} & 9.04$\times 10^8$ & 42.44$\times 10^8$ \\
\hline
BCDU-Net \cite{bcdunet} & 20.66$\times 10^8$ & 39.76$\times 10^8$\\
\hline
Attn U-Net \cite{attentionunet} & 8.91$\times 10^8$ & 17.82$\times 10^8$\\
\hline
FocusNet \cite{focusnet} & 19.07$\times 10^8$ & 91.36$\times 10^8$\\
\hline
FocusNet++ & \textbf{7.80$\times 10^8$} & \textbf{15.64$\times 10^8$}\\
\hline
\end{tabular}
\end{center}
\caption{Comparing the model complexity and performance (on ISIC 2018) for FocusNet++ against state of the art segmentation architectures.}
\label{isiccomplexity}
\end{table}

The ISIC 2018 skin cancer segmentation dataset \cite{isic2018} has become a major benchmark dataset for the evaluation of medical imaging algorithms. We use the 2594 images with corresponding ground truths for our experiments. We divide these images into a training set of 1815 images, a validation set of 259 images, and a test set of 520 images. We resize every images to a smaller $256\times256$ size, via an anti-aliasing down-sampling technique.\\
Table \ref{isic2018falpha} summarizes our results for the experiments. FocusNet++ significantly outperforms every architecture across all metrics for the ISIC 2018 dataset with considerably fewer parameters and FLOPs. We get a 4.6\% higher JI over the next best model. Table \ref{isiccomplexity} summarizes the number of parameters and FLOPs of each architecture.

\subsection{Cell Nuclei Segmentation}
\label{cellnuclei}

We now consider the segmentation of smaller regions inside images. For this, we use the cell nuclei segmentation dataset \cite{kaggle}, which was a part of the Data Science Bowl 2018. It contains 670 images which we divided into a training set of 540 and a validation set of 130. We resize all images to $256\times256$. For this task, we evaluate the performance of our architecture by reducing the number of parameters (via reducing the number of filters per layer) for it in a way that it has fewer than one million FLOPs. We also reduced the number of parameters for the other architectures to account for the smaller size of this dataset so that we don't overfit. Our results are summarized in Table \ref{nucleisegmentationap}. FocusNet++ outperforms BCDU-Net with 2.5 times fewer parameters and 10 times fewer FLOPs.

\begin{table}[htb]
\begin{center}
\begin{tabular}{|c||c|c|c|c|c|c|}
\hline
Method & Params & FLOPs & Precision & Recall\\
\hline
\hline
U-Net \cite{unet} & 3.62 & 1.89  & 0.8976 & 0.9052 \\
\hline
BCU-Net \cite{bcdunet} & 5.22 & 9.98 & 0.9024 & 0.9078\\
\hline
Attn U-Net \cite{attentionunet} & 2.32 & 1.84 & 0.8782 & 0.9019\\
\hline
FocusNet \cite{focusnet} & 5.03 & 22.38 & 0.9016 & 0.8981\\
\hline
FocusNet++ & \textbf{1.84} & \textbf{0.98} & \textbf{0.9173} & \textbf{0.9139}\\
\hline
\end{tabular}
\end{center}
\caption{Segmentation results on the cell nuclei segmentation dataset. Params and FLOPs are of the order of $\times 10^8$.}
\label{nucleisegmentationap}
\end{table}

\section{Conclusion}
\label{conclusion}

We proposed an extremely efficient and accurate medical image segmentation architecture, FocusNet++, based on our novel residual group attention block that outperforms existing state-of-the-art architectures. We also propose an extremely lightweight variant of this architecture that outperforms architectures that are almost 2.5 times its size. We adapt the Tversky loss and balanced cross entropy loss in the adaptive logarithmic loss setting to boost performance over true positives and true negatives in order to obtain more well-rounded segmentations. Based on our experiments, our architecture requires lesser parameters and FLOPs, while giving better results compared to other architectures.

\vfill
\pagebreak

\section{Compliance with Ethical Standards}

This research study was conducted retrospectively using human subject data made available in open access by \cite{isic2018} and \cite{kaggle}. Ethical approval was not required as confirmed by the license attached with the open access data.

\section{Acknowledgements}

Chaitanya Kaul and Roderick Murray-Smith acknowledge support from the iCAIRD project, funded by Innovate UK (project number 104690. No other conflicts of interest.

\bibliographystyle{IEEEbib}
\bibliography{refs}

\end{document}